\documentclass[letterpaper,11pt]{article}
\usepackage{tabularx} 
\usepackage{amsmath}  
\usepackage{graphicx} 
\usepackage{amsmath}
\usepackage{floatrow}
\usepackage{placeins}
\usepackage[utf8]{inputenc}
\usepackage[T1]{fontenc}
\usepackage{authblk}
\usepackage{soul}
\usepackage{dblfloatfix}
\usepackage{algorithm}
\usepackage{algpseudocode}

\usepackage[justification=centering]{caption}
\usepackage{caption}
\usepackage{color}

\usepackage[margin=1in,letterpaper]{geometry} 
\usepackage[final]{hyperref} 
\hypersetup{
	colorlinks=true,       
	linkcolor=blue,        
	citecolor=blue,        
	filecolor=magenta,     
	urlcolor=blue         
}

\begin{document}
\newcommand\tab[1][0.5cm]{\hspace*{#1}}

\title{Fluid-structure interaction simulations for the prediction of fractional flow reserve in pediatric patients with anomalous aortic origin of a coronary artery}
\author[1]{Charles Puelz}
\author[1]{Craig G. Rusin}
\author[1]{Dan Lior}
\author[1]{Shagun Sachdeva}
\author[1]{Tam T. Doan}
\author[1]{Lindsay F. Eilers}
\author[1]{Dana Reaves-O'Neal}
\author[1]{Silvana Molossi}
\affil[1]{Department of Pediatrics, Division of Cardiology, Baylor College of Medicine and Texas Children's Hospital}


\maketitle

\begin{abstract}
Computer simulations of blood flow in patients with anomalous aortic origin of a coronary artery (AAOCA) have the promise to provide insight into this complex disease. They provide an {\em in-silico} experimental platform to explore possible mechanisms of myocardial ischemia, a potentially deadly complication for patients with this defect. This paper focuses on the question of model calibration for fluid-structure interaction models of pediatric AAOCA patients. Imaging and cardiac catheterization data provide partial information for model construction and calibration. However, parameters for downstream boundary conditions needed for these models are difficult to estimate. Further, important model predictions, like fractional flow reserve (FFR), are sensitive to these parameters. We describe an approach to calibrate downstream boundary condition parameters to clinical measurements of resting FFR. The calibrated models are then used to predict FFR at stress, an invasively measured quantity that can be used in the clinical evaluation of these patients. We find reasonable agreement between the model predicted and clinically measured FFR at stress, indicating the credibility of this modeling framework for predicting hemodynamics of pediatric AAOCA patients. This approach could lead to important clinical applications since it may serve as a tool for risk stratifying children with AAOCA.
\end{abstract}

\section{Introduction}

Anomalous aortic origin of a coronary artery (AAOCA) is one of the leading causes of sudden cardiac death in the young, yet it may have no clinical expression in many patients \cite{Molossi20}. Risk stratification in AAOCA remains difficult because abnormal morphologic features can be seen in patients with and without myocardial ischemia. Thus, it is unclear which patients may benefit from surgical intervention and which are safe to continue to exercise, especially in the setting of right (R-)AAOCA, a subtype 4-6 times more prevalent than left (L-)AAOCA \cite{Molossi24, Doan23a, Doan23b}. Invasive intracoronary flow assessment via cardiac catheterization may be helpful in determining risk in those patients with concerning clinical symptoms and no evidence of inducible myocardial ischemia in non-invasive studies under stress conditions (exercise or pharmacological) \cite{Agrawal21, Doan21, Doan20}. This study seeks to determine the feasibility of non-invasive simulations to predict intracoronary flow and pressure in AAOCA, thus mimicking results obtained during a cardiac catheterization procedure.

There have been numerous computational studies of coronary physiology, including AAOCA. These studies range from stand-alone compartmental descriptions to fluid-structure interaction (FSI) models that rely on compartmental models for downstream boundary conditions. For normal coronary physiology, Arthurs et al.~created a compartmental simulator that included a sophisticated description of control under exercise \cite{Arthurs16}. As part of their work, they coupled this coronary model to flow in a three-dimensional aortic geometry. With respect to models of AAOCA physiology, Cesarani et al.~developed a process for creating patient-specific compartmental models of AAOCA that rely on imaging and catheterization data \cite{Ceserani23}. With the models created in their study, they demonstrated good agreement between model predicted and clinically measured coronary flow. Razavi et al.~created image-based computational fluid dynamics simulations and used them to study shear stresses in the anomalous coronary and other clinically relevant quantities \cite{Razavi21, Razavi22}. Complementary to these fluid mechanics studies, Formato et al.~and Rosato et al.~simulated the deformation of the vessel walls within parametrized and image-based AAOCA anatomies \cite{Formato18, Rosato23}. Finally, Jiang et al.~constructed detailed FSI simulations that described deformation of the aortic and coronary walls coupled to the blood flow \cite{Jiang22}. They demonstrated reasonable agreement between model predicted and clinically measured values at rest and in pharmacologically induced stress.

The modeling approach in this paper extends the work of Jiang et al., with several important differences worth highlighting \cite{Jiang22}. First, we consider models of pediatric patients, instead of adults, as well as patients with both L-AAOCA and R-AAOCA. Second, our image-based geometries incorporate an anisotropic vessel wall thickness based on the local radius of curvature. This approach allows for realistically thinner coronary walls compared to the thicker aortic wall. Most significantly, parameters for the coronary boundary conditions are determined in a new way that is patient-specific and based only on the clinically measured resting FFR. The main question we seek to address is whether FFR during stress conditions, an important clinical indicator of high-risk physiology and compromised intracoronary flow, can be accurately predicted using models calibrated only with the resting FFR. Such a result would support the credibility of this framework for predicting clinically relevant hemodynamics outside of the resting scenario in which they are calibrated.

\section{Methods}

\subsection{Clinical data and model construction}

Computed tomography angiography (CTA) images and cardiac catheterization data were gathered from 10 pediatric AAOCA patients (3 L-AAOCA, 7 R-AAOCA) seen at the Heart Center at Texas Children's Hospital. This data was collected under protocols H-32955 and H-43968 approved by the Baylor College of Medicine Institutional Review Board. Our standardized approach for clinical evaluation has been previously published \cite{Molossi20, Molossi24}, and a cardiac catheterization study is performed on a small subset of patients with conflicting clinical data, as described in the previous section.  Catheterization data are measured at rest and during pharmacological stress induced from dobutamine and adenosine \cite{Agrawal21}. Fractional flow reserve (FFR), defined as the ratio of mean coronary pressure to mean aortic pressure, is computed from pressure waveforms measured in the anomalous coronary vessel \cite{Doan20}. FFR serves as a measure of coronary flow, in order to detect obstruction and thus severity of disease. An FFR value close to 1 indicates a non-stenotic vessel, while a value less than 1 indicates some level of obstruction to flow and possible risk of myocardial ischemia. 

\begin{figure}[h!]
    \centering
    \includegraphics[scale=0.45]{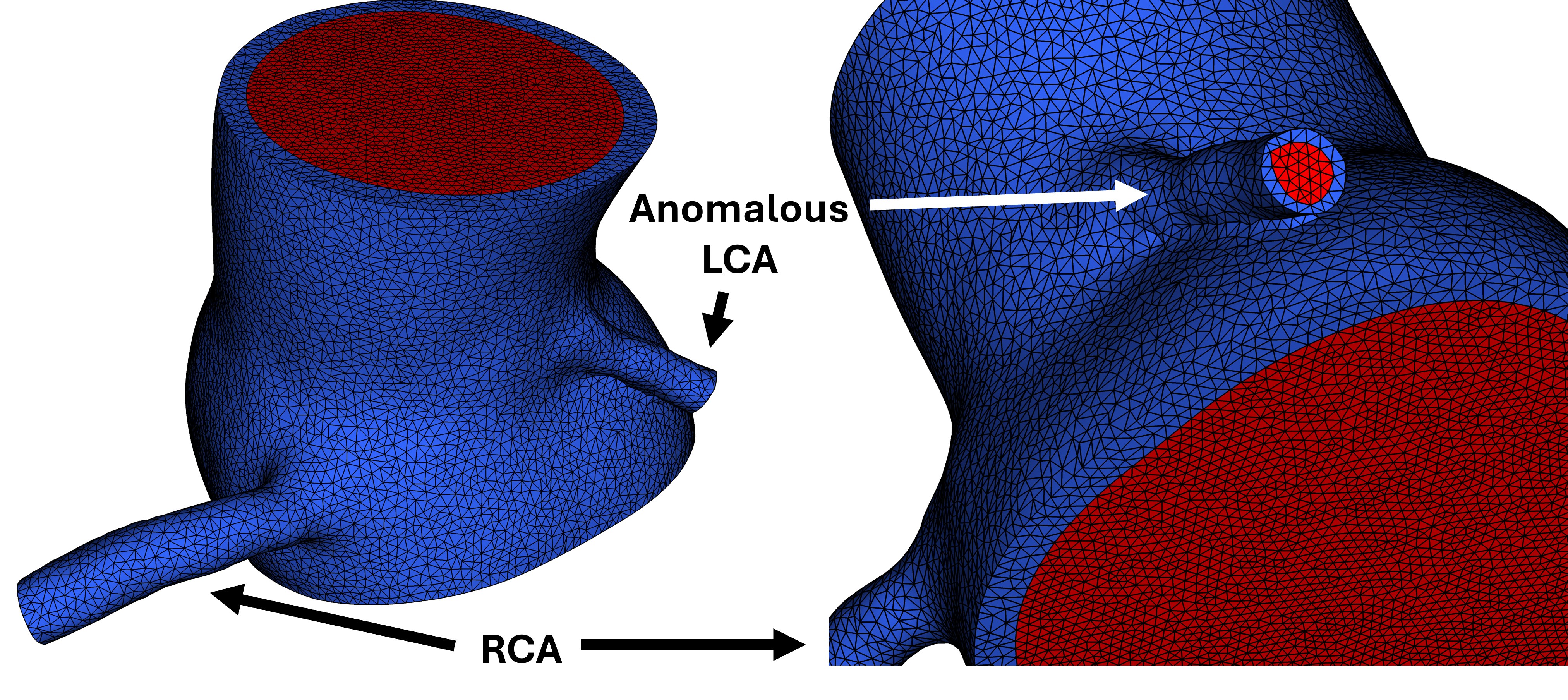}
    \caption{A visualization of the finite element meshes created for a patient with an anomalous left coronary artery. The fluid mesh is colored red and the wall mesh is colored blue. Two different perspectives are shown. The wall mesh includes the intramural segment of the anomalous left coronary artery. Note that the thickness of the coronary wall is smaller than the thickness of the aortic wall and is based on the local radius of curvature of the segmented lumen surface.}
    \label{fig:mesh}
\end{figure}

Finite element meshes for the lumen and vessel wall are needed for the FSI simulations. Aortic meshes that include the left (LCA) and right (RCA) coronary arteries are constructed. Mesh construction follows three steps: segmentation, postprocessing, and tetrahedral mesh generation. For each patient, a segmentation of the aorta and main coronaries is created in 3D Slicer from the corresponding CTA data \cite{Kikinis2013}. This segmentation is imported into MeshMixer for postprocessing. The geometry is trimmed to include only the two proximal coronary arteries (namely, the normal origin coronary artery and the anomalous one) as well as regions above and below the sinotubular junction. In order to define the aortic and coronary vessel walls, a smoothed version of the segmented surface is extruded in the direction of the area weighted normal vector by an amount that depends on the local radius of curvature. Care is taken in this extrusion step to define the {\em intramural segment} for the anomalous coronary, largely considered a high-risk feature of this disease. This segment is present in many AAOCA patients and defines a region over which the aorta and anomalous coronary share part of their vessel walls. Deformation of this segment is thought to be a cause of coronary flow obstruction, especially during the high pressures induced from stress/exercise.  Note that the extrusion process, based on the local radius of curvature, results in realistic vessel wall thicknesses, with a thicker aortic wall and thinner coronary wall (refer to Figure \ref{fig:mesh}).  Separate closed surface meshes are then created for the vessel wall and the lumen. The surface shared by the wall and lumen meshes is comprised of the same triangles due to the requirement of conforming fluid and solid meshes in the numerical method. These surface meshes are then imported into SimVascular. Tetrahedral meshes are generated and sidesets for the meshes are labeled for the imposition of boundary conditions.

\subsection{Fluid-structure interaction modeling}

FSI simulations are executed within the SimVascular software framework, which has proven useful in modeling blood flow and tissue deformation in adult R-AAOCA patients \cite{Jiang22, Updegrove17}. Similarly to what was done by Jiang et al., FSI simulations are initialized with a pre-stressed vessel wall \cite{Jiang22}. The at-rest diastolic pressure measured in the catheterization lab is used to define the traction on the lumen wall needed for the prestress calculation. The elasticity modulus for the wall is set to $1.5 \times 10^7 \text{ dynes} \cdot \text{cm}^{-2}$ for all patients \cite{Jiang22}.  The pre-stressed vessel wall, along with zero flow and diastolic pressure initial conditions in the vessel lumen, are used to initialize the FSI simulations. Seven to eight cardiac cycles are simulated, with four at rest and three to four at stress. Systolic and diastolic pressures at rest and stress along with the rest and stress cardiac cycle lengths are used to define boundary conditions for the simulations, to be described below. 

\subsection{Boundary conditions}
\label{subsec:bcs}

\begin{figure}[h!]
    \centering
    \includegraphics{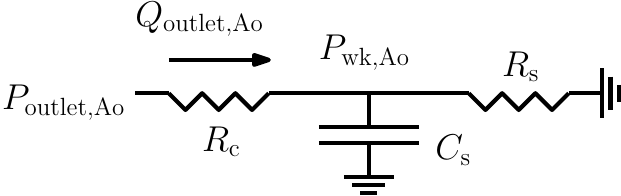}
    \caption{Schematic of the three-element Windkessel model used to describe the peripheral vasculature downstream of the aorta in the fluid-structure interaction model.}
    \label{fig:aooutlet}
\end{figure}

A pressure boundary condition is specified at the inlet of the ascending aorta. The minimum and maximum pressures $P_\text{dias}$ and $P_\text{sys}$ and the cardiac cycle length $T$ are set to values measured in the catheterization lab. An analytic form for the aortic pressure is specified and based on these three clinically measured parameters:
\begin{align}
    P_\text{inlet,Ao}(t) = 
    \begin{cases}
        \frac{(P_\text{sys} - P_\text{dias})}{\exp(-k_\text{Ao})}\, \exp\left(\frac{-k_\text{Ao}}{1 - g(t)^2}\right) + P_\text{dias} & 0 \leq t < t_\text{valve} \\
        \gamma\, \exp(-\beta t) & t_\text{valve} \leq t < T 
    \end{cases},
    \label{eq:inletAo}
\end{align}
where $t_\text{peak} = 0.3\, T$, $t_\text{valve} = t_\text{peak} + 0.05\, T$, $k_\text{Ao} = 2$, $g(t) = \frac{2\, t}{2\, t_\text{peak}} - 1$, and 
\begin{align}
    \tilde{P} &= \frac{(P_\text{sys} - P_\text{dias})}{\exp(-k_\text{Ao})}\, \exp\left(\frac{-k_\text{Ao}}{1 - g(t_\text{valve})^2}\right) + P_\text{dias} ,\\ 
    \beta &= \frac{1}{T - t_\text{valve}}\log\left(\frac{\tilde{P}}{P_\text{dias}}\right),\\
    \gamma &= \frac{\tilde{P}}{\exp(-\beta\, t)}.
\end{align}
The pressure at the outlet of the ascending aorta, denoted $P_\text{outlet,Ao}$, is determined from a three-element Windkessel model with parameters $R_\text{c}$, the characteristic resistance, $R_\text{s}$, the systemic vascular resistance, and $C_\text{s}$, the systemic vascular compliance. Refer to Figure \ref{fig:aooutlet}.

The three element Windkessel model has the characteristic resistance in series with a parallel circuit comprised of the systemic vascular resistance and compliance. The Windkessel pressure $P_\text{wk,Ao}$ downstream from the characteristic resistance satisfies the following equation:
\begin{align}
    \frac{d}{dt}(C_\text{s} P_\text{wk,Ao}) = Q_\text{outlet,Ao} - \frac{P_\text{wk,Ao}}{R_\text{s}},
\end{align}
with $Q_\text{outlet,Ao}$ equal to the flow at the outlet of the aorta. The outlet pressure boundary condition is determined from the following equation:
\begin{align}
    P_\text{outlet,Ao} = P_\text{wk,Ao} + \frac{Q_\text{outlet,Ao}}{R_\text{c}}.
\end{align}

\begin{figure}[h!]
    \centering
    \includegraphics{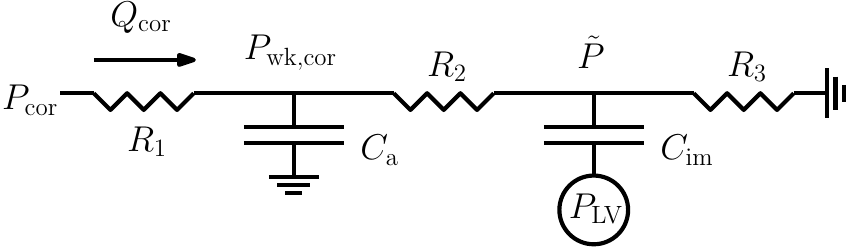}
    \caption{Schematic of the five-element Windkessel model used to describe the peripheral vasculature downstream of the coronaries in the fluid-structure interaction model.}
    \label{fig:cor_outlet}
\end{figure}

Pressure boundary conditions for the coronary arteries, denoted $P_\text{cor}$, are determined from a Windkessel model with five elements: $C_\text{a}$, $C_\text{im}$, $R_1$, $R_2$, and $R_3$. Calibration of the total effective resistance for the coronary model is described in subsection \ref{subsec:calibration}. This model is similar to the three element Windkessel, except for the inclusion of an additional $RC$ circuit in which the capacitor is not grounded. Instead, it is coupled to the left ventricular pressure $P_\text{LV}$ to simulate the effect of ventricular contraction and relaxation on coronary blood flow. Refer to Figure \ref{fig:cor_outlet}.

The equations for the Windkessel pressure $P_\text{wk,cor}$ and additional pressure $\tilde{P}$ are
\begin{align}
    \frac{d}{dt}(C_\text{a} P_\text{wk,cor}) &= Q_\text{cor} + \frac{\tilde{P} - P_\text{wk,cor}}{R_\text{2}}, \label{eq:cor1} \\
    \frac{d}{dt}(C_\text{im} (\tilde{P} - P_\text{LV})) &= \frac{P_\text{wk,cor} - \tilde{P}}{R_\text{2}} - \frac{\tilde{P}}{R_3}, \label{eq:cor2}
\end{align}
where $Q_\text{cor}$ is the flow into the coronary. The coronary pressure boundary condition is then determined by the following equation:
\begin{align}
    P_\text{cor} = P_\text{wk,cor} + \frac{Q_\text{cor}}{R_1}.
\end{align}
The left ventricular pressure is specified analytically. Its form depends on the systolic pressure and the period of the cardiac cycle:
\begin{align}
    P_\text{LV}(t) = 
    \begin{cases}
        0 & 0 \leq t < t_\text{shift}\\
         \frac{P_\text{sys}}{\exp(-k_\text{vent})}\, \exp\left(\frac{-k_\text{vent}}{1 - h(t)^2}\right) & t_\text{shift} \leq t < t_\text{sys} \\
        0 & t_\text{valve} < t < T 
    \end{cases},
\end{align}
where $t_\text{shift} = 0.0692\, T$, $t_\text{s} = 0.4\, T$, $k_\text{vent} = 0.3$, and $h(t) = \frac{2\, (t - t_\text{shift})}{t_\text{s} - t_\text{shift}} - 1$. This function is differentiated analytically and used in equation \eqref{eq:cor2}. Note that only the derivative of this pressure waveform appears in the model, so the pertinent information is the difference in minimum and maximum ventricular pressures, determined by $P_\text{sys}$, and not their individual values.

In order to describe changes in the resistances and compliances in stress, a simple model is used that maintains the $RC$ time constant while encoding the known drop in the vascular resistance. More precisely, resistances in each Windkessel model are halved and compliances are doubled for all patients. The stress model for the peripheral vasculature, downstream of the aorta, is:
\begin{align}
    R_\text{s} &\rightarrow R_\text{s} / 2, \\
    C_\text{s} &\rightarrow 2 \times C_\text{s},
\end{align}
and the stress model for the coronary vasculature is:
\begin{align}
    R_\text{2} &\rightarrow R_\text{2} / 2, \\
    C_\text{a} &\rightarrow 2 \times C_\text{a}, \\
    C_\text{im} &\rightarrow 2 \times C_\text{im}.
\end{align}
Systolic and diastolic pressures and cardiac cycle lengths are taken from the cardiac catheterization data at both rest and stress.

\subsection{Parameter selection and calibration}
\label{subsec:calibration}

Patient specific boundary condition parameters are determined from catheterization data and the literature. For the three element Windkessel model at the aortic outlet, we use standard formulas to approximate the systemic vascular resistance and compliance in the resting state. The systemic vascular resistance is computed as: 
\begin{align}
    R_\text{s} = \frac{\frac{2}{3}P_\text{dias} + \frac{1}{3}P_\text{sys}}{\text{cardiac output}},
\end{align}
and the systemic vascular compliance is computed as:
\begin{align}
    C_\text{s} = \frac{\text{stroke volume}}{P_\text{sys} - P_\text{dias}}.
\end{align}
The characteristic resistance for this model is set to $R_\text{c} = 10$ dynes $\cdot$ s $\cdot$ cm$^{-5}$ for all patients. 

\begin{figure}[h!]
    \centering
    \includegraphics[width=\textwidth]{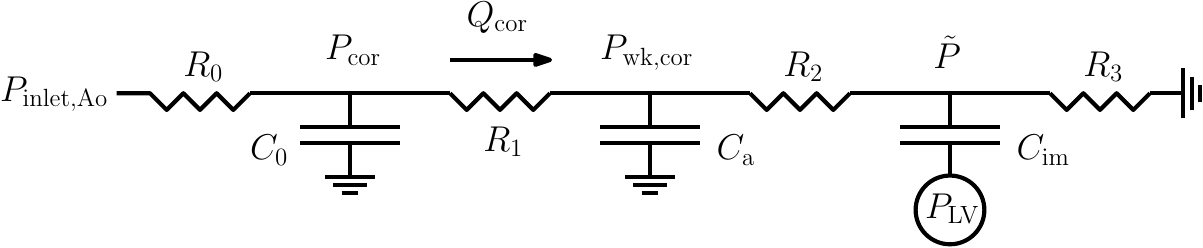}
    \caption{Schematic of the surrogate Windkessel model used to calibrate the total effective resistance of the five-element coronary Windkessel model.}
    \label{fig:surrogate}
\end{figure}

For the coronary model, two sets of resistance parameters are considered: a nominal set that is used for all patients and a set that is patient-specific and calibrated to the resting FFR data. First, nominal parameters for the three resistors are set to $R_1 = 3.1992 \times 10^4$ dynes $\cdot$ s $\cdot$ cm$^{-5}$, $R_2 = 6.3984 \times 10^4$ dynes $\cdot$ s $\cdot$ cm$^{-5}$, $R_3 = 2.666 \times 10^5$ dynes $\cdot$ s $\cdot$ cm$^{-5}$ based on previous modeling studies \cite{Arthurs16}. The calibrated set of parameters is computed in several steps, which is summarized in Algorithm \ref{alg:cal}. The two compliance parameters in the coronary Windkessel model are set to $C_\text{a} = 4.5 \times 10^{-7}$ cm$^5$/dynes and $C_\text{im} = 3 \times 10^{-6}$ cm$^5$/dynes for all patients based on previous studies \cite{Arthurs16}. Next, the total resistance of the Windkessel model for the anomalous coronary is calibrated using a stand-alone and computationally efficient surrogate model. This surrogate model is driven by the resting inlet aortic pressure specified in equation \eqref{eq:inletAo}. A schematic of the surrogate model is shown in Figure \ref{fig:surrogate}. 

To drive the stand-alone surrogate model with this pressure, an additional $RC$ circuit with parameters $R_0$ and $C_0$ is needed. This pair of compartments is placed upstream of the five element coronary Windkessel model. The resistance and compliance of the additional circuit define an additional set of compartments that approximates the FSI part of the full model. We assume the compliance of this compartment is relatively small and of the same order of magnitude as $C_\text{a}$, so $C_0 = 1\times 10^{-7}$ cm$^5$/dynes. The resistance $R_0$ is determined by running an additional FSI simulation for each patient. The vessel wall is pre-stressed to a 60 mmHg pressure load and then a pressure gradient across the coronary is linearly applied at a rate of 37.5 mmHg/s. The pressure gradient as a function of the flow through the anomalous coronary is recorded, and $R_0$ is set to the derivative of this curve evaluated at a pressure gradient of 40 mmHg. Finally, the total effective resistance of the anomalous coronary Windkessel model is calculated by computing a scale factor $\alpha$ for the nominal resistances $R_1$, $R_2$, and $R_3$, using the surrogate model, so that the model-predicted rest FFR is equal to the rest FFR value measured in the catheterization lab. Since the mean pressures behave monotonically with the resistances, the calibration procedure is implemented by placing the surrogate model within a bisection method to compute the scale factor. 

\begin{algorithm}
\caption{Calibration of the coronary boundary conditions for the FSI model.}\label{alg:cal}
\begin{algorithmic}[1]
\Require Nominal coronary resistances $R_1$, $R_2$, and $R_3$ (see Figure \ref{fig:cor_outlet}).
\State Run an FSI simulation to estimate $R_0$ for the surrogate model (see Figure \ref{fig:surrogate}).
\State Compute $\alpha$ so the surrogate model with $R_0$, $\alpha R_1$, $\alpha R_2$, and $\alpha R_3$ predicts the rest FFR.
\State Run an FSI simulation in rest and stress with calibrated resistances $\alpha R_1$, $\alpha R_2$, and $\alpha R_3$.
\Ensure Predictions of FFR in rest and stress from the FSI simulation.
\end{algorithmic}
\end{algorithm}

\section{Results}

\begin{figure}[h!]
    \centering
    \includegraphics[scale=0.35]{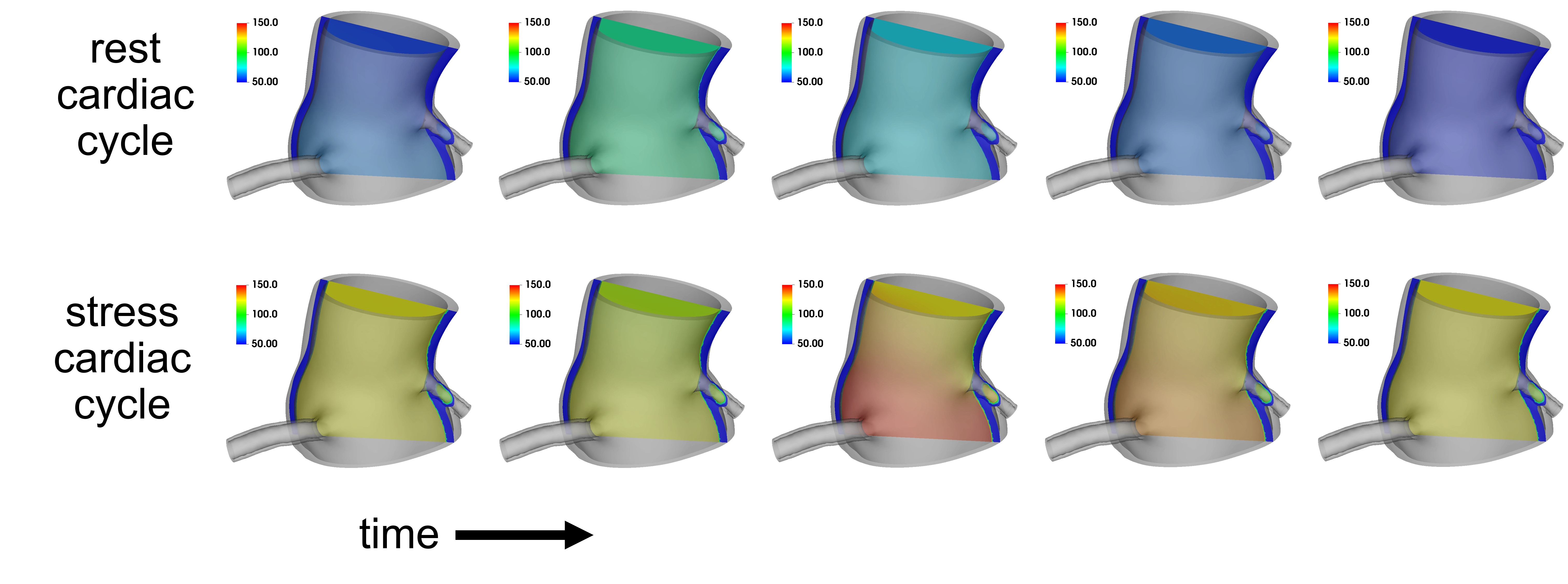}
    \caption{A visualization of simulation results for cardiac cycles at rest and in stress. The wall mesh is displayed in translucent grey. A slice of the pressure field is shown that includes both the aortic lumen and the lumen of the proximal anomalous left coronary artery.}
    \label{fig:rest_and_stress}
\end{figure}

\begin{figure}[h!]
    \centering
    \includegraphics[scale=0.3]{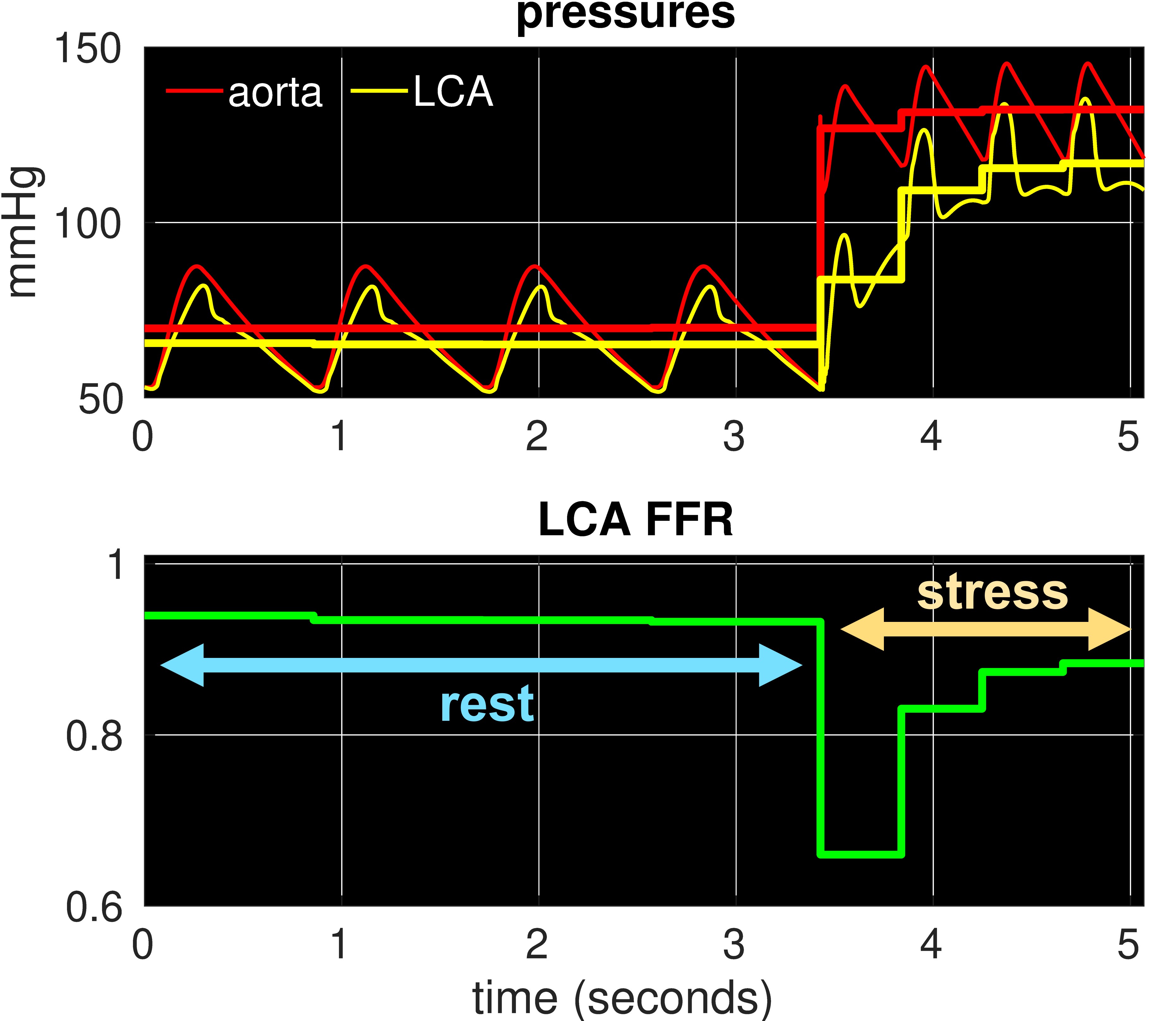}
    \caption{The top panel shows pressure waveforms in the aorta and anomalous left coronary artery and the bottom panel shows the calculated FFR. There are four cycles at rest followed by four cycles at stress. Notice the drop in FFR in stress, similar to what is seen in the catheterization lab.}
    \label{fig:volcano}
\end{figure}

In this section, we describe results from our FSI models constructed from the data of 10 pediatric AAOCA patients in our study. Results from a simulation of an L-AAOCA patient are shown in Figure \ref{fig:rest_and_stress}. The top and bottom rows contain snapshots in time from one rest and one stress cardiac cycle, respectively. The deforming vessel wall is shown in translucent grey along with a slice of the pressure field that includes both the aortic lumen and anomalous coronary lumen. Note the difference in pressure between the anomalous coronary and the aorta during stress. The model is able to capture this clinically observed pressure gradient, which is likely caused in part by the deforming intramural segment inherent to this anatomy. It would not be possible to describe this mechanism using a purely fluid mechanics simulation with stationary vessel walls. Figure \ref{fig:volcano} is an example of the pressure and FFR waveforms produced by our simulations. The top panel shows the aortic and proximal intracoronary pressures for a patient with L-AAOCA. The bottom panel shows the FFR values computed from these pressure waveforms for each cardiac cycle. For this patient, four cycles were simulated at both rest and stress. Here, we can clearly see a larger difference between the aorta and LCA pressures in stress, leading to a drop in FFR.

\begin{figure}[h!]
    \centering
    \includegraphics[scale=0.35,trim=0 0 -50 0]{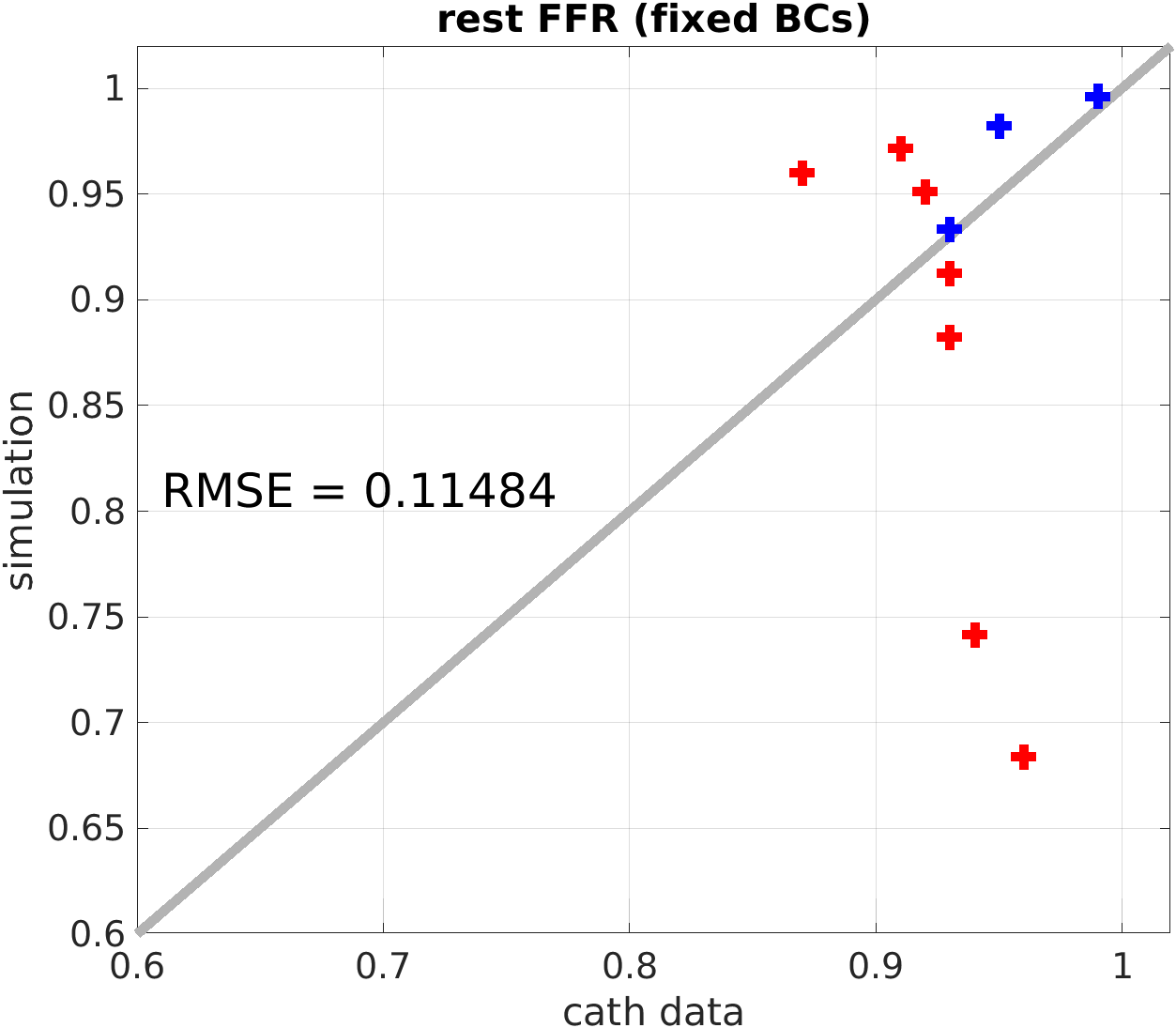}
    \includegraphics[scale=0.35,trim=0 0 0 0]{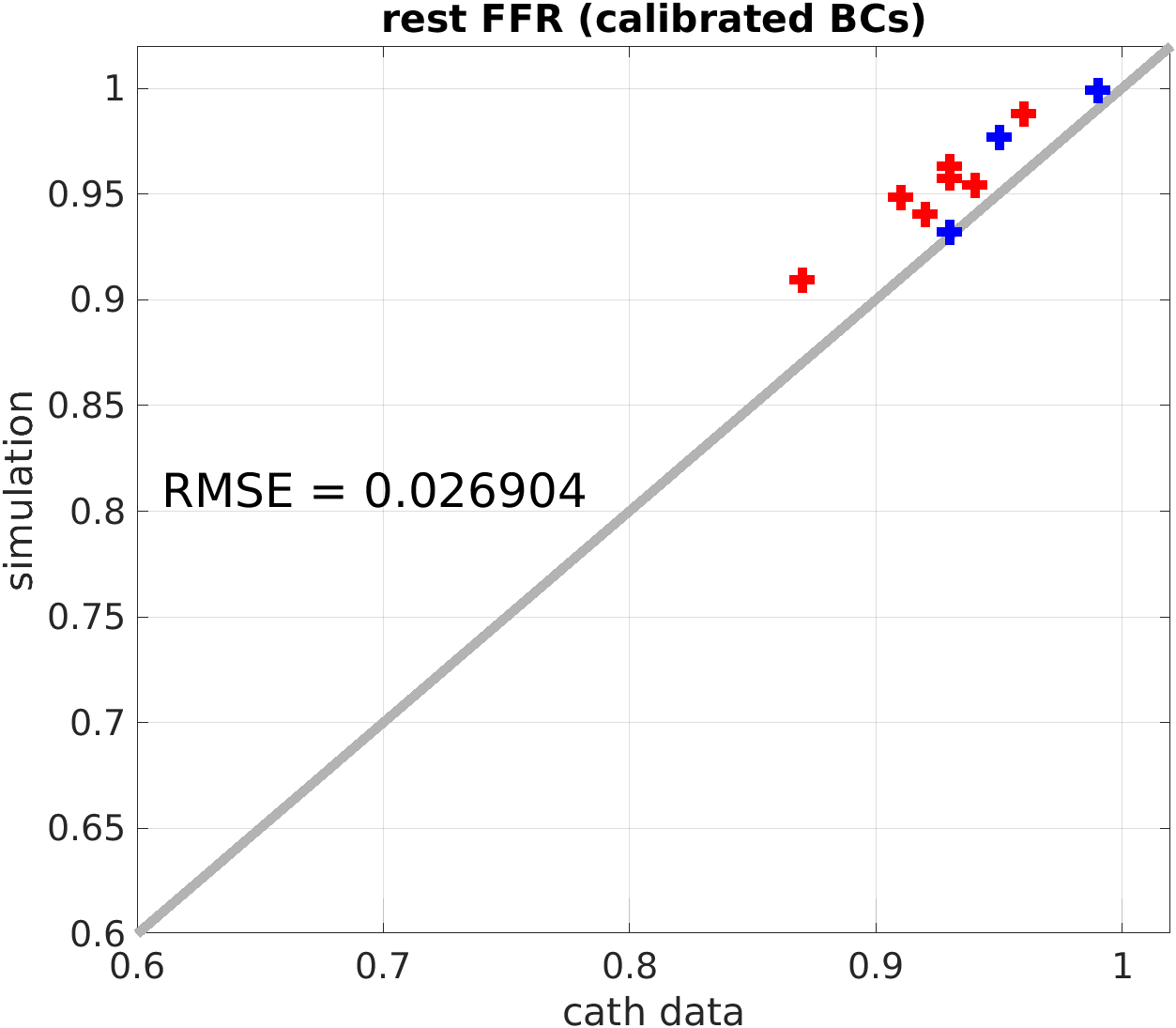} \\
    \includegraphics[scale=0.35,trim=0 0 -50 -50]{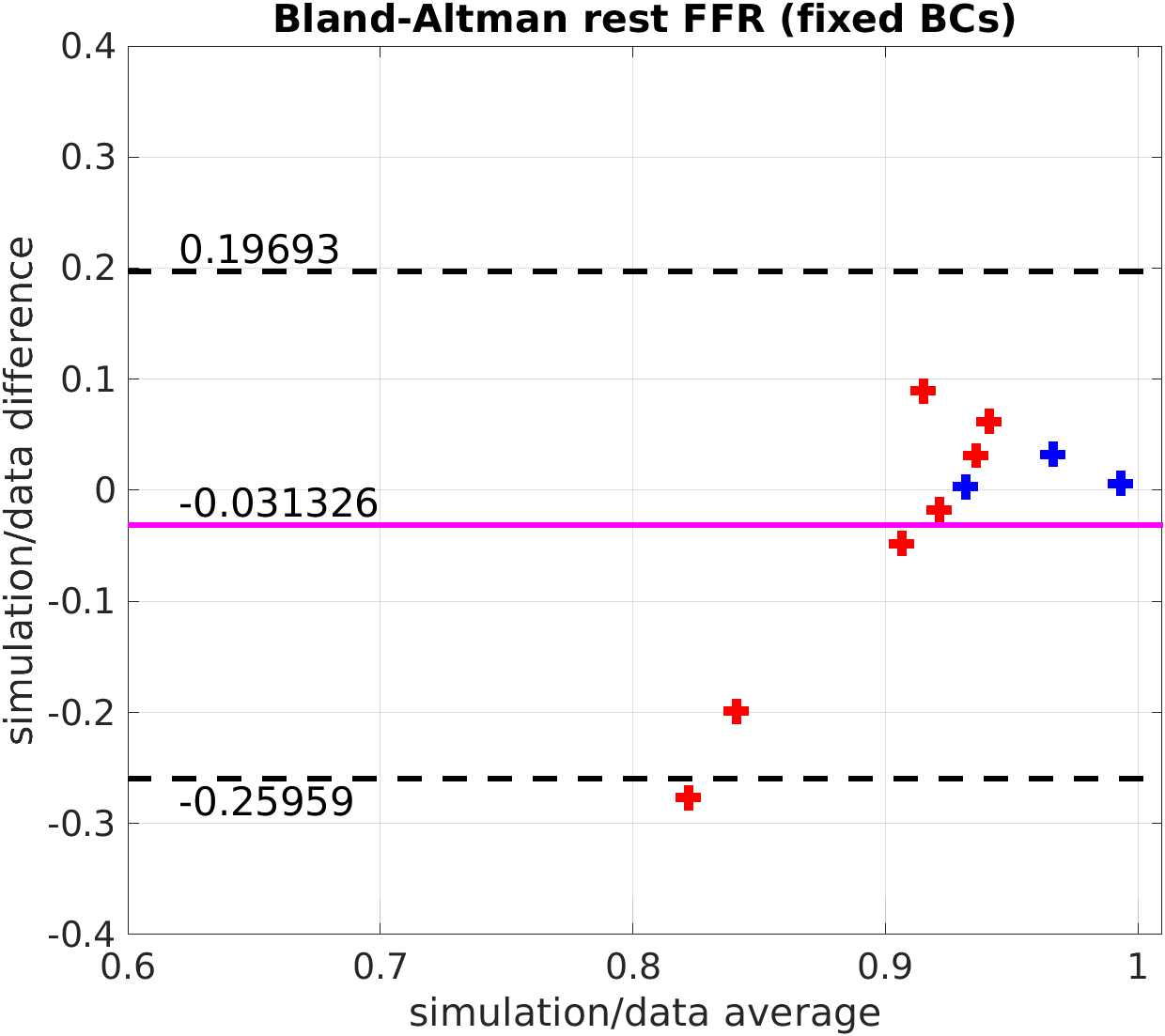}
    \includegraphics[scale=0.35,trim=0 0 0 -50]{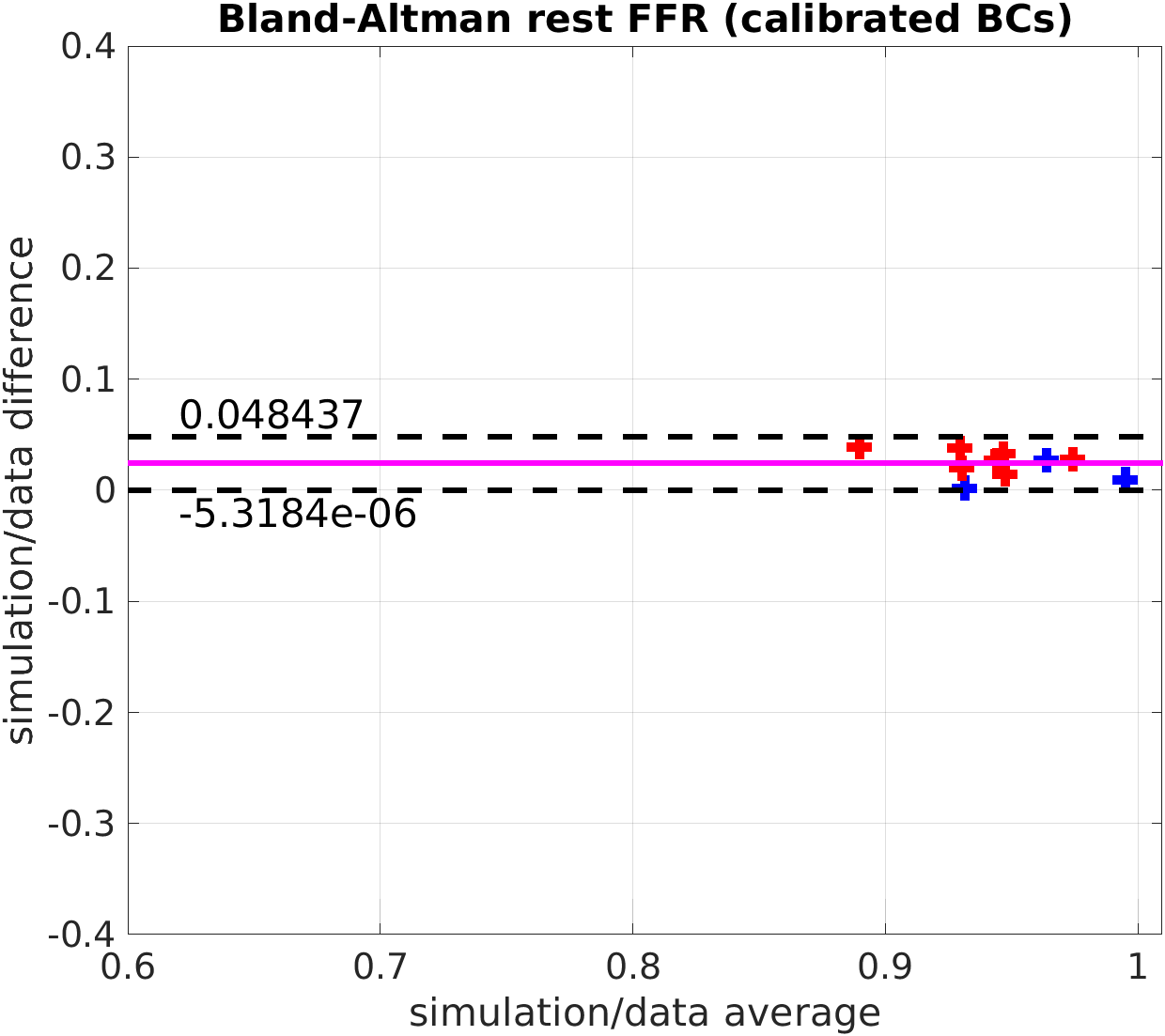}
    \caption{Bland-Altman plots for the rest FFR. The left panel corresponds to simulations with fixed coronary boundary condition parameters for all patients. The right panel corresponds to simulations with coronary boundary condition parameters that were calibrated to the clinically measured rest FFR. Red and blue corresponds to R- and L-AAOCA, respectively.}
    \label{fig:ffr_baseline}
\end{figure}

To see the effect of the calibration procedure for the coronary boundary conditions described in subsection \ref{subsec:calibration}, results are shown for two sets of coronary boundary condition parameters: (i) a nominal, or fixed set of parameters values used for all patients, based on Arthurs et al. \cite{Arthurs16}, and (ii) calibrated sets of parameter values that are specific for each patient and determined by Algorithm \ref{alg:cal}. The calibration process depends on the rest FFR, and Figure \ref{fig:ffr_baseline} depicts parity and Bland-Altman plots for the predicted rest FFR \cite{Bland86}. The left plots correspond to the fixed coronary boundary conditions and the right plots corresponds to the calibrated boundary conditions. Blue and red colors correspond to L- and R-AAOCA respectively. The calibrated boundary conditions result in a substantially reduced root-mean-square-error (RMSE) between the predicted and clinically measured rest FFR values, as compared to the fixed boundary conditions. However, the model predicted rest FFR values are not exactly the same as the measured values because the surrogate model used to calibrate the boundary conditions is only an approximation of the full FSI model.

\begin{figure}[h!]
    \centering
    \includegraphics[scale=0.35,trim=0 0 -50 0]{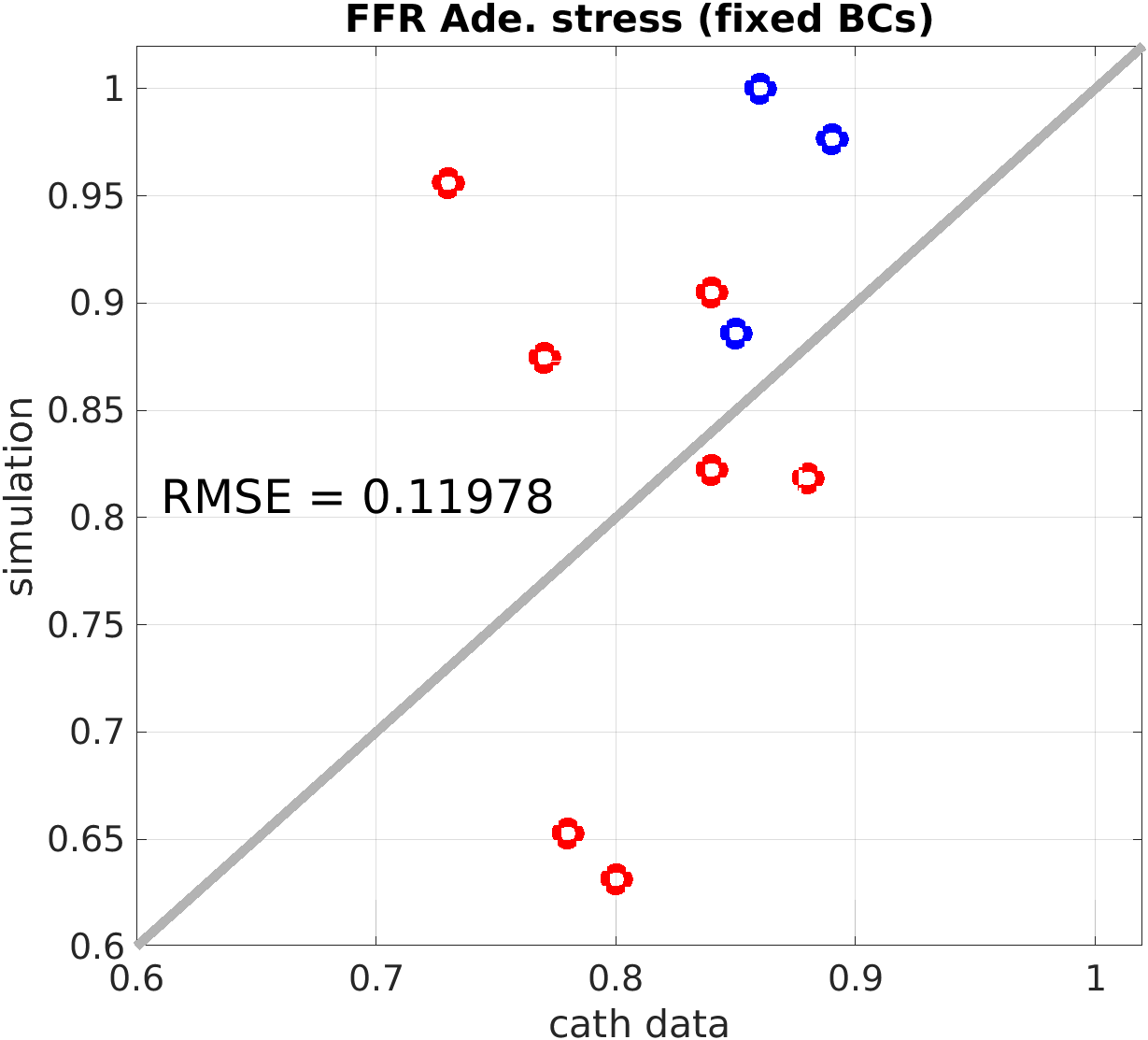}
    \includegraphics[scale=0.35,trim=0 0 0 0]{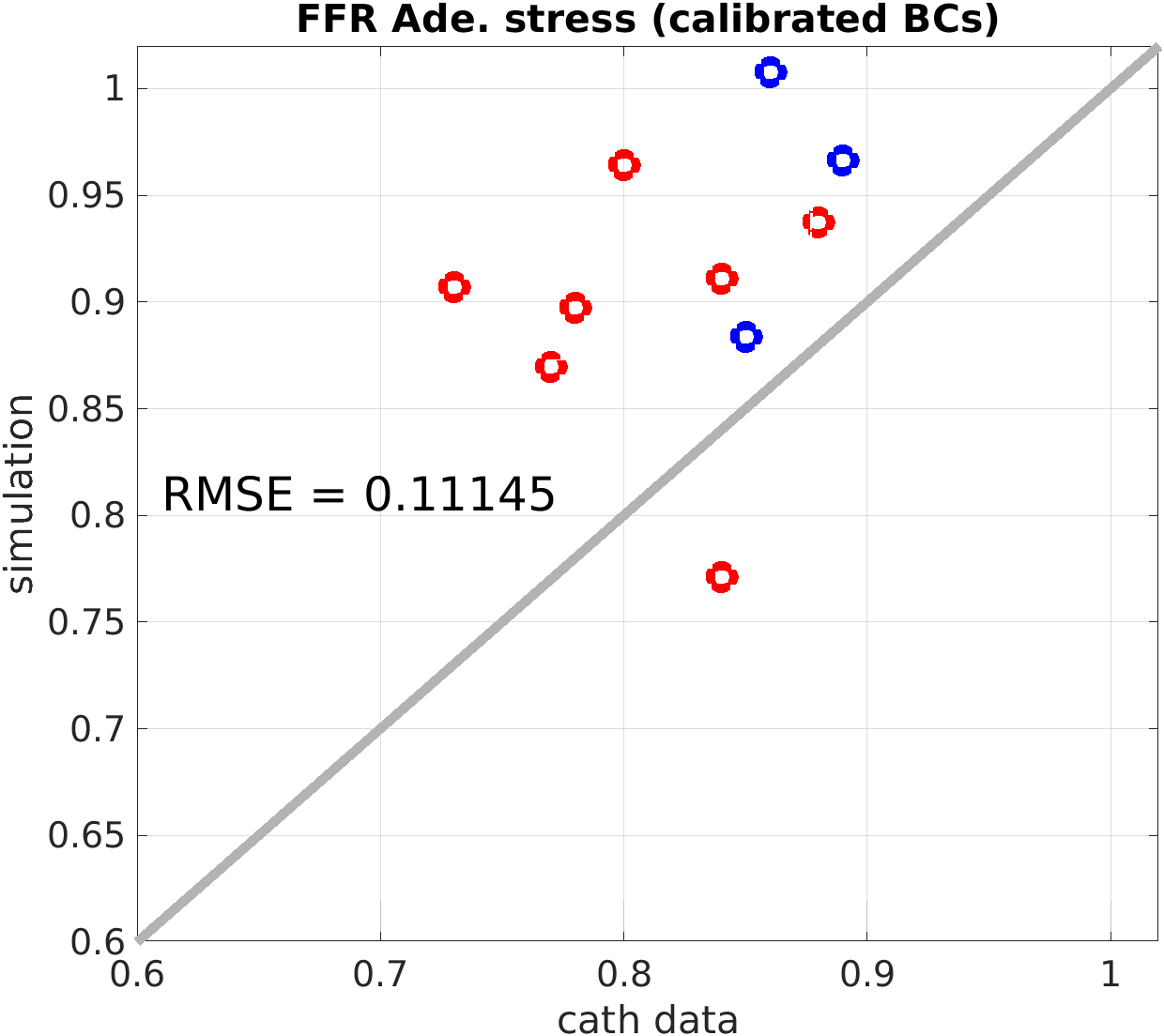} \\
    \includegraphics[scale=0.35,trim=0 0 -50 -50]{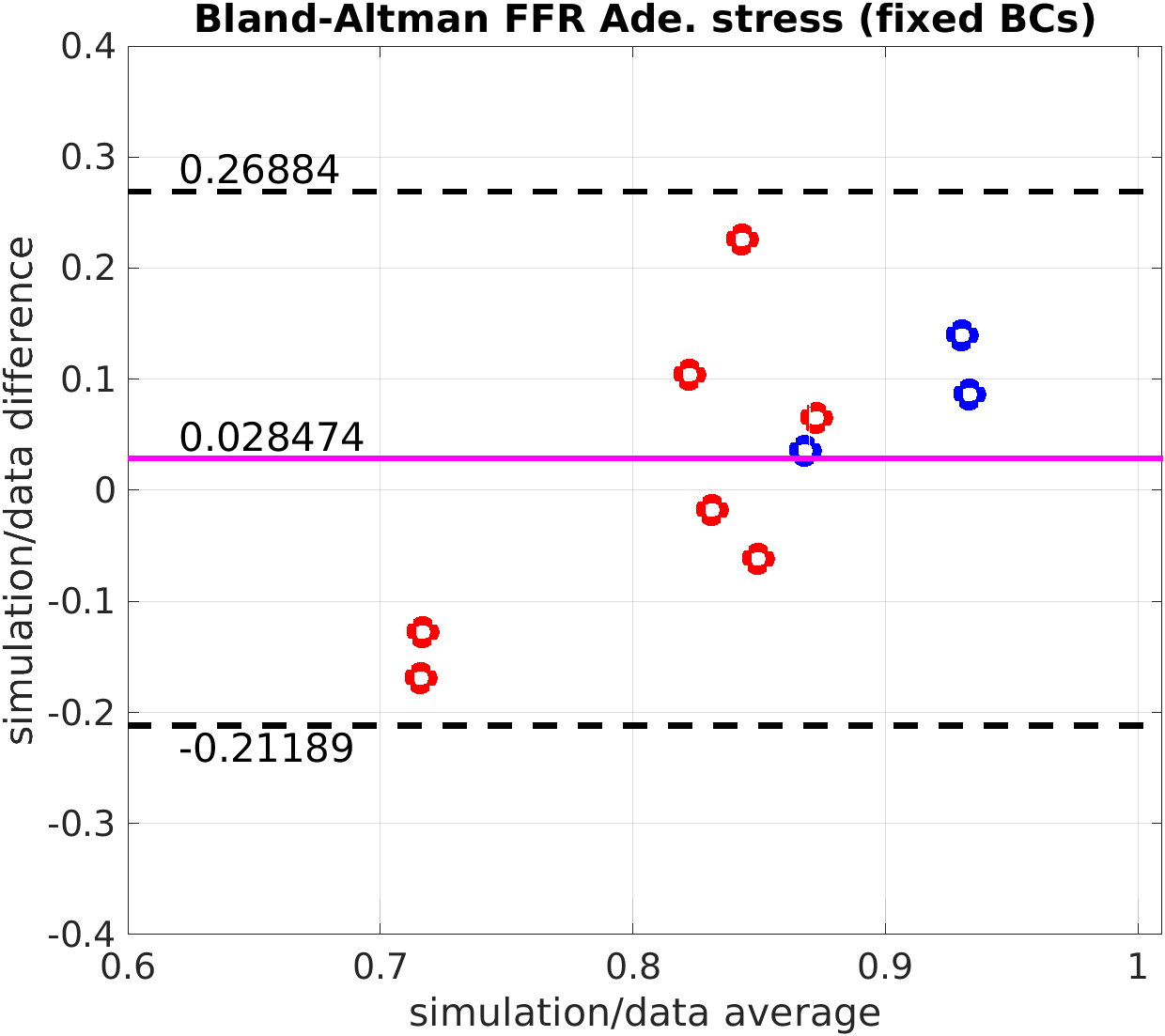}
    \includegraphics[scale=0.35, trim=0 0 0 -50]{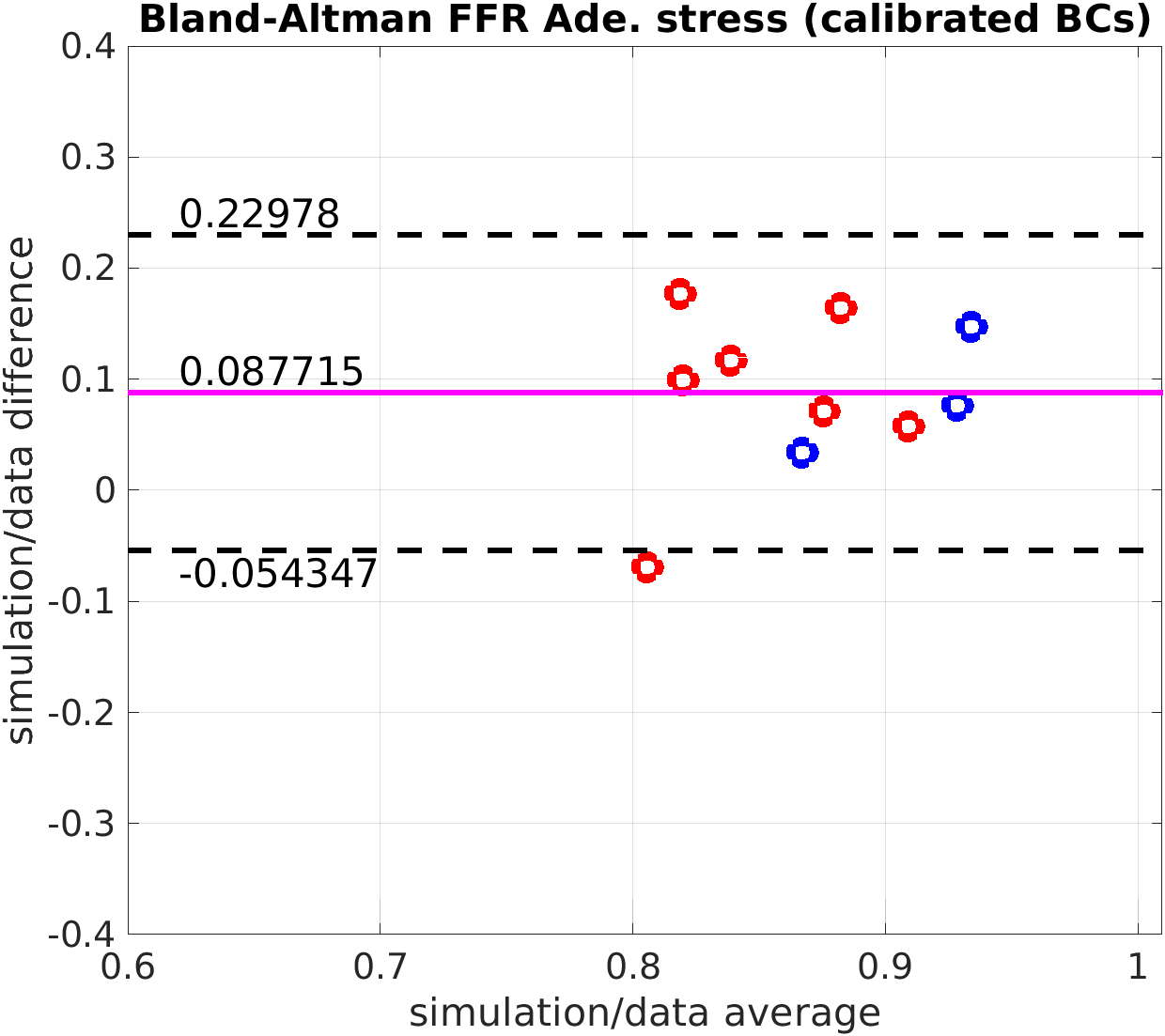}
    \caption{The top panels are parity plots for FFR under adenosine stress. The left panel corresponds to simulations with fixed coronary boundary condition parameters for all patients. The right panel corresponds to simulations with coronary boundary condition parameters that were calibrated to the clinically measured rest FFR. Red and blue corresponds to R- and L-AAOCA, respectively. The bottom panels are the corresponding Bland-Altman plots for the stress FFR.}
    \label{fig:ffr_ade}
\end{figure}

\begin{figure}[h!]
    \centering
    \includegraphics[scale=0.35,trim=0 0 -50 0]{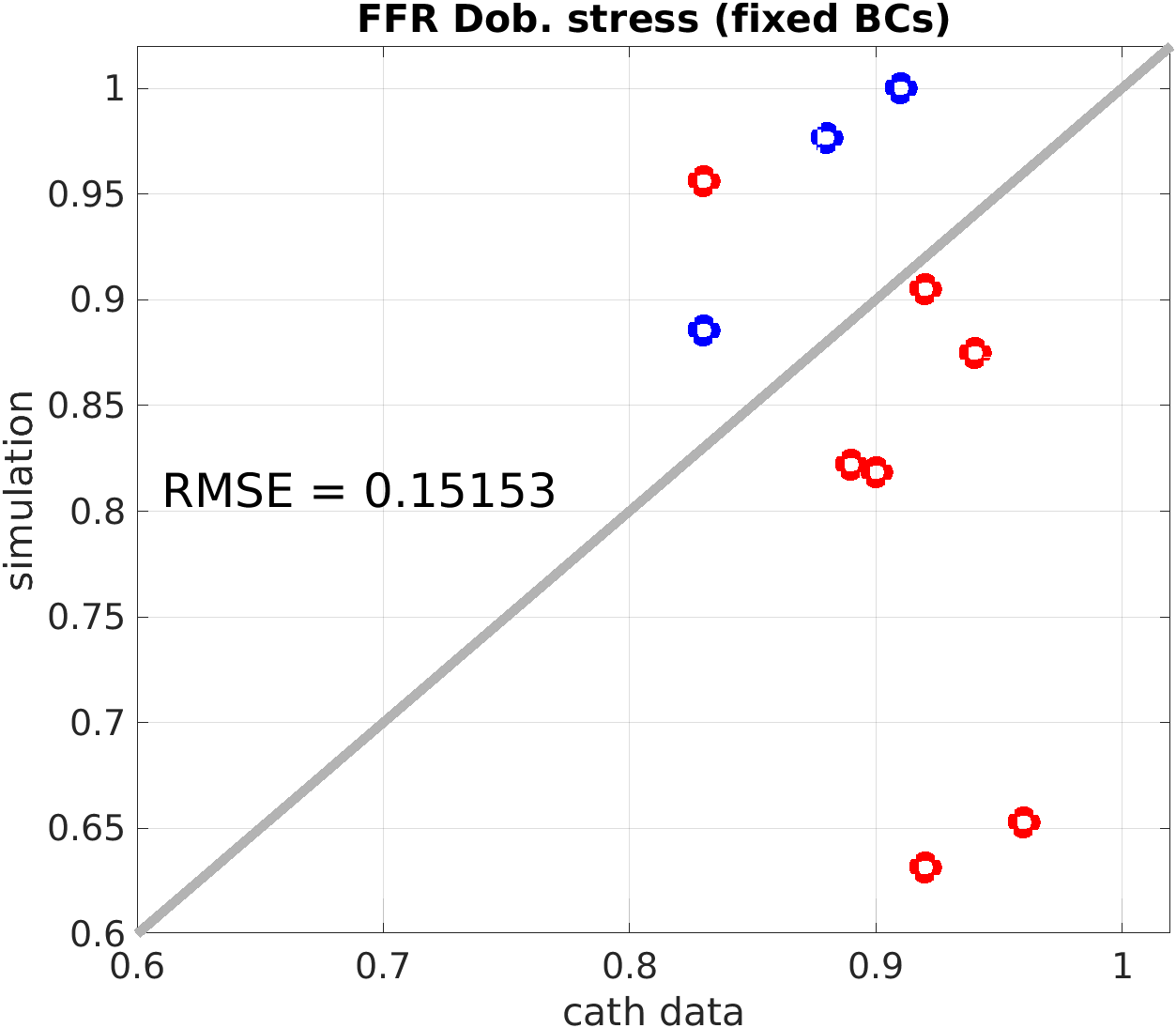}
    \includegraphics[scale=0.35]{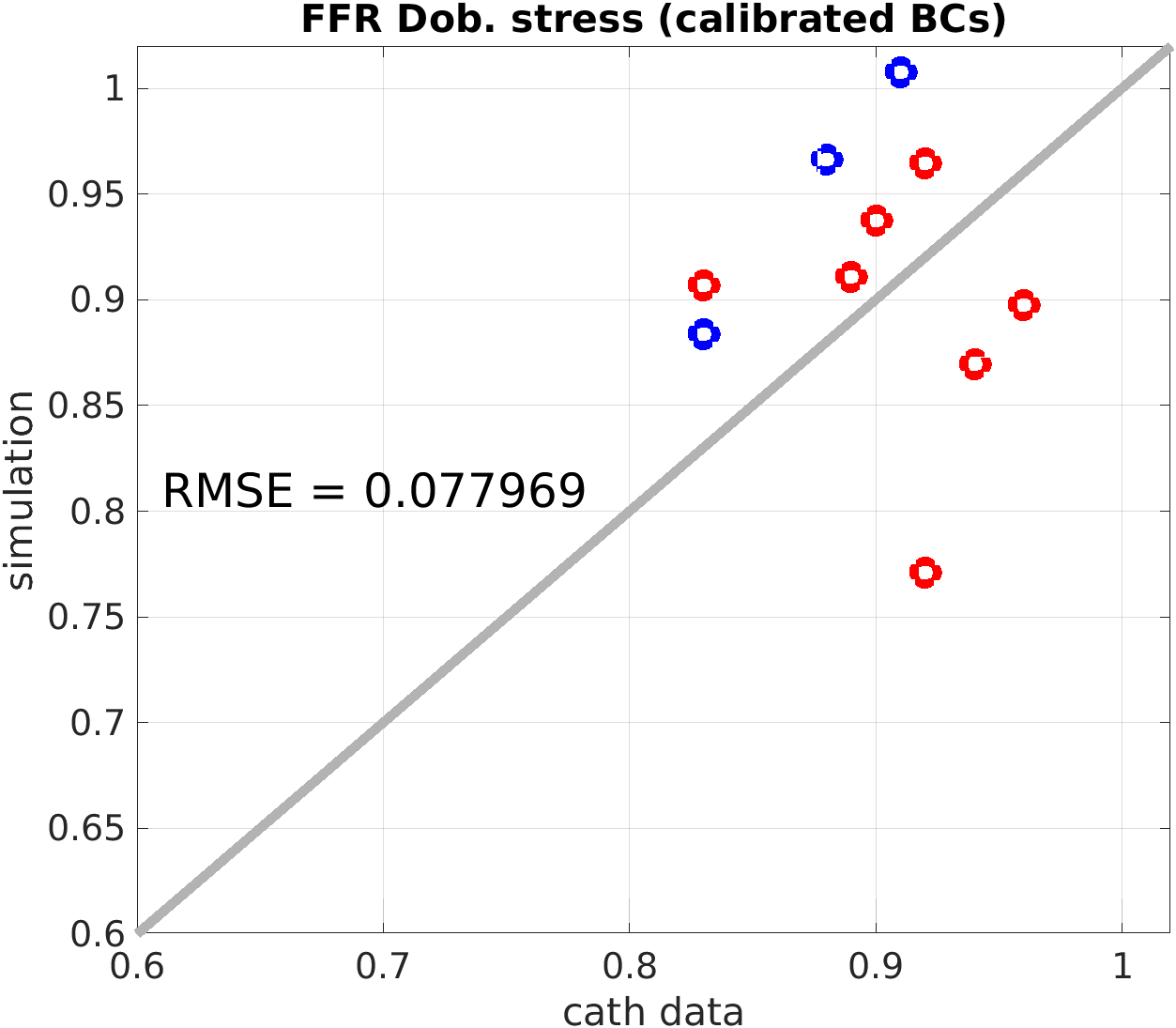} \\
    \includegraphics[scale=0.35,trim=0 0 -50 -50]{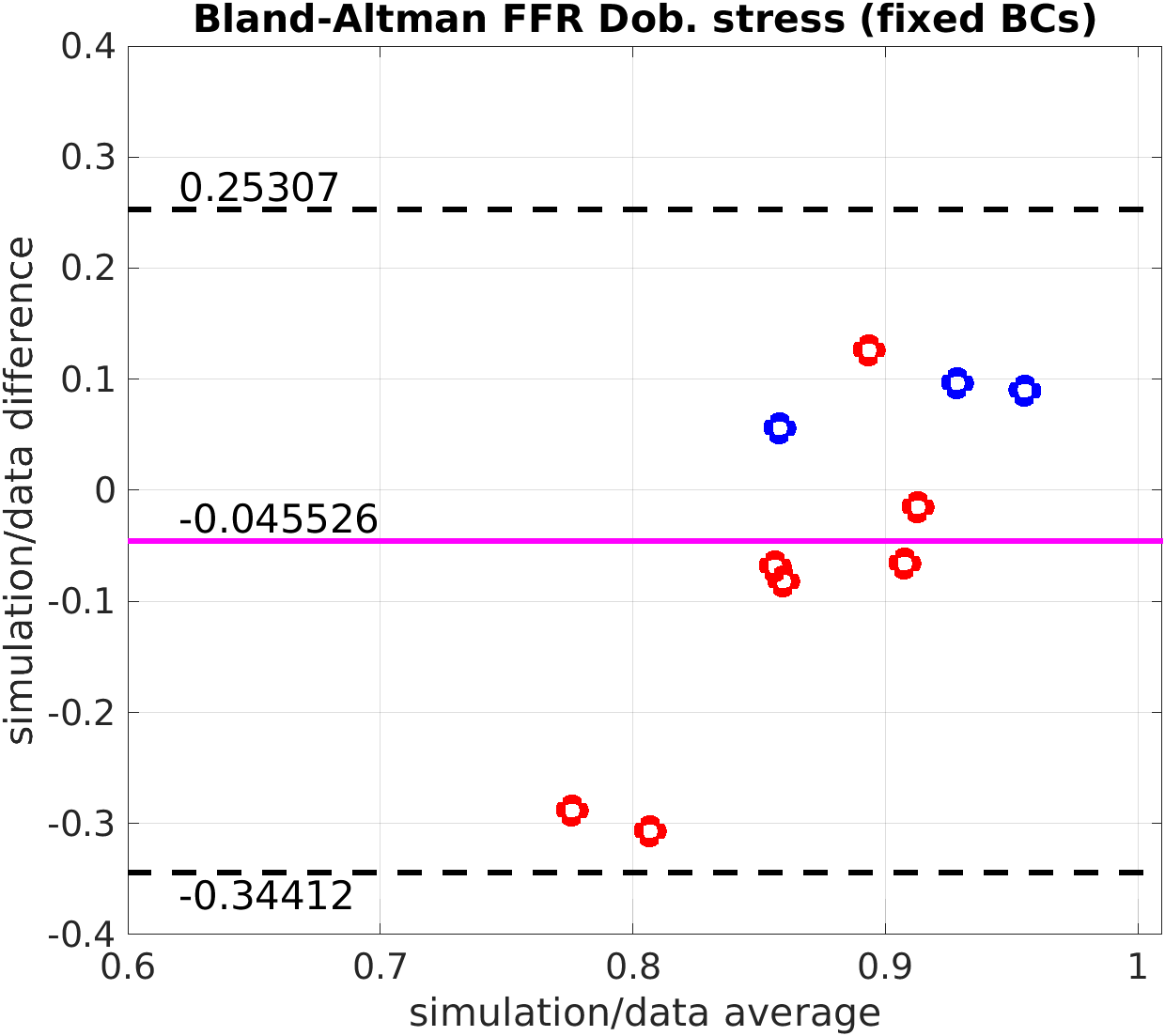}
    \includegraphics[scale=0.35, trim=0 0 0 -50]{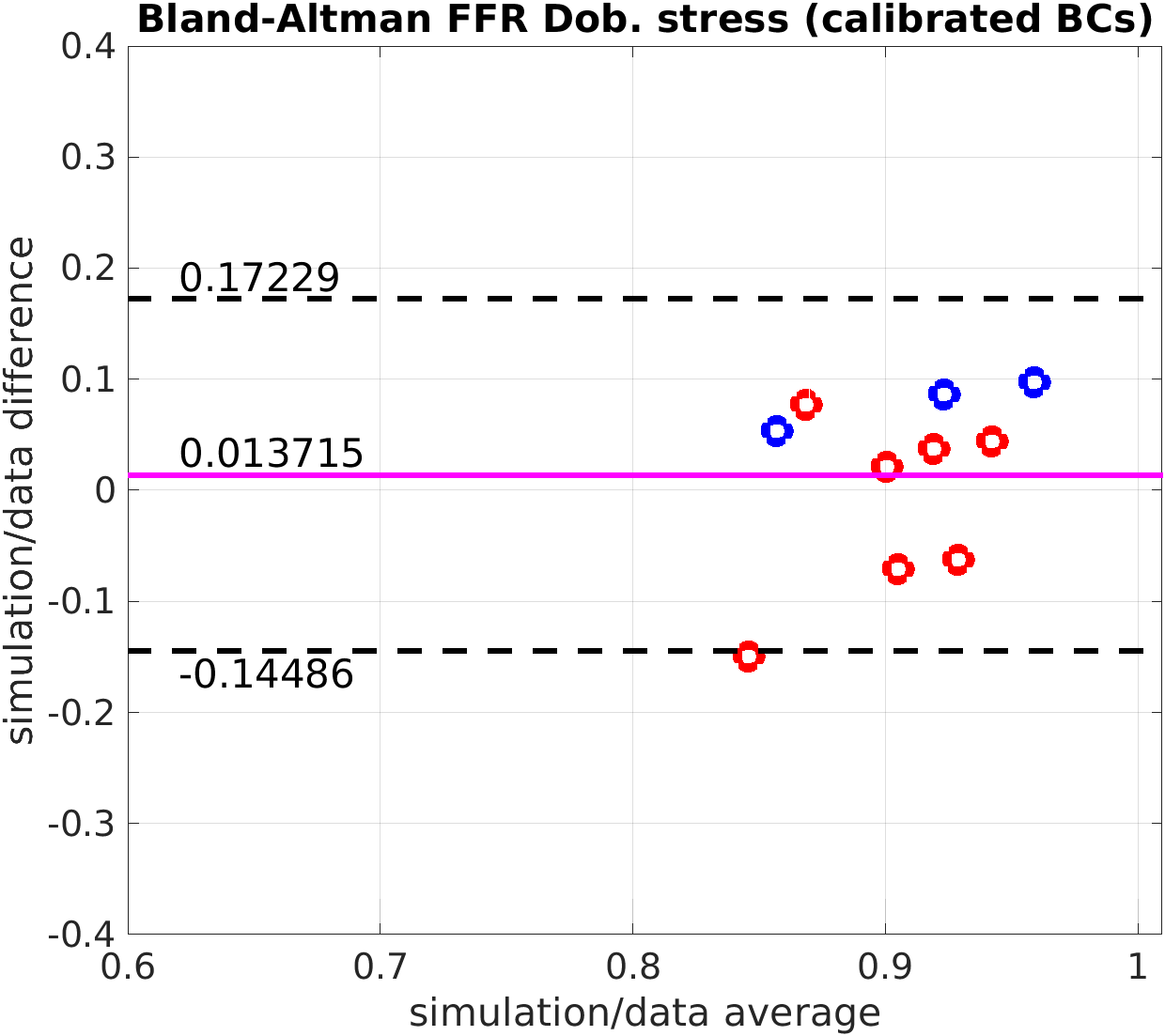}
    \caption{The top panels are parity plots for FFR under dobutamine stress. The left panel corresponds to simulations with fixed coronary boundary condition parameters for all patients. The right panel corresponds to simulations with coronary boundary condition parameters that were calibrated to the clinically measured rest FFR. Red and blue corresponds to R- and L-AAOCA, respectively. The bottom panels are the corresponding Bland-Altman plots for the stress FFR.}
    \label{fig:ffr_dob}
\end{figure}

Figures \ref{fig:ffr_ade} and \ref{fig:ffr_dob} show parity and Bland-Altman plots for the scenarios of adenosine and dobutamine induced stress. Our model for the transition from rest to stress, described in subsection \ref{subsec:bcs}, does not distinguish between adenosine and dobutamine, so we compare results from our simulations to clinical data from each of these scenarios. 

When the simulations are compared to the adenosine stress FFR values, there is only a small, essentially negligible improvement in the RMSE between the fixed and calibrated boundary conditions. The Bland-Altman plots seen in Figure \ref{fig:ffr_ade} indicate an increase in the bias from fixed to calibrated boundary conditions. However, the variance in the error does decrease.

The decrease in the RMSE, between the fixed and calibrated boundary conditions, is more dramatic for the case of dobutamine stress. There is approximately a 50\% improvement in the RMSE from the fixed case to calibrated case. The Bland-Altman plots in Figure \ref{fig:ffr_dob} indicate both a substantial decrease in bias as well as a decrease in the variance in the discrepancy between the simulation and clinical data. 

It is not immediately clear why the model predictions align better with the measurements in dobutamine stress. A clue might lie in imaging studies in animals and humans that compare stress induced from adenosine and dobutamine \cite{Lafitte01, Paetsch04}. These studies found differences in the stress physiologic response between these two drugs. In general, dobutamine better identified lower grade lesions. Furthermore, we are of the opinion (at our center) that dobutamine is able to identify potential fixed and dynamic components of flow obstruction associated with the anomalous coronary in stress. Both types of mechanisms are naturally captured in the FSI model. Nonetheless, further work is needed to understand differences between adenosine and dobutamine induced stress and the relationship of the physiologic responses of these drugs to the stress model used in this paper.   

\clearpage 

\section{Discussion and Conclusions}

In this paper, we described an approach for computer modeling of pediatric AAOCA. Our main goal was to see if FSI models could reliably predict stress FFR, since this number is used in clinically challenging cases to assess disease severity and aid in risk stratification for management decision-making. Computer models were constructed using CTA imaging data for patient-specific geometries and cardiac catheterization data for patient-specific boundary conditions. We considered two scenarios for coronary boundary conditions: those that were fixed across all patients and those that were calibrated for each patient using their clinically measured resting FFR. Calibration of the coronary boundary conditions involved a multistep procedure with a simplified surrogate model. This approach was necessary because the FSI simulations were too complex and computationally expensive to allow for the direct calibration of their boundary conditions.

First, we found that our calibration process resulted in dramatically improved predictions of rest FFR. However, these predictions for rest FFR were not exact. This is to be expected, since the computationally efficient surrogate model, used for the calibration procedure, is only an approximation of the FSI model. Second, we found that the models with calibrated boundary conditions predicted stress FFR with more accuracy than the models with fixed boundary conditions, especially when compared to stress induced from dobutamine. This finding demonstrated that coronary boundary conditions can be calibrated at rest to predict FFR in stress, and that the fluid-structure interaction framework presented in this study was able to capture various mechanisms leading to reduced FFR in stress with reasonable fidelity.

There are several limitations and areas for improvement with respect to the modeling framework presented here. Of note is the simplified nature for the model describing changes in the downstream coronary vasculature in stress. The type of stress model used in this study was chosen purposefully for a few reasons. First, a main goal was to compare predictions in stress FFR across all patients, and we did not want to confound these results with complicated patient-specific stress models. Second, if we chose to calibrate a patient-specific model for stress dynamics, it is not clear how to estimate parameters for such a model using only data in the resting state, such as FFR. Finally, dynamic changes in stress and exercise are extremely difficult to characterize mechanistically, especially within a computer modeling framework and for pediatric patients with complex disease. Such technical and modeling issues are certainly of interest, but they are outside of the scope of the current study. Another limitation is the simplicity of the surrogate model and the way in which its parameters were chosen. The surrogate model used here replaces the FSI part of the model (ascending aorta and main coronaries) with a single $RC$ circuit in series with the existing coronary boundary condition model. Conceivably, this surrogate model could be made more complex, either by adding additional compartments or by representing the aorta and coronaries with a network of one-dimensional vessels. However, it is unclear how to determine the additional parameters associated with these more complex models. Our surrogate model, with the addition of a single $RC$ circuit, required only one additional fluid-structure interaction simulation to determine the single resistance parameter.  

Fluid-structure interaction models that accurately predict FFR during stress conditions could transform the way that AAOCA patients are clinically evaluated, especially if these approaches could enhance and/or provide additional insight beyond invasive catheterization procedures that can be of high risk. This work supports the credibility of such models for predicting stress FFR. However, additional research is necessary to both improve these models as well as to integrate them into the clinical treatment and workflow for patients living with AAOCA.


\section{Acknowledgements}

This work was supported in part by the Big-Data Private-Cloud Research Cyberinfrastructure MRI-award funded by NSF under grant CNS-1338099 and by Rice University's Center for Research Computing.

\bibliographystyle{plain}
\bibliography{references}

\end{document}